\documentclass[amsmath,amssymb,nofootinbib]{revtex4}

\pdfoutput=1

\usepackage{graphicx}
\usepackage{dcolumn}
\usepackage{bm}
\usepackage{epsfig}
\usepackage{color}
\usepackage{wasysym} 

\usepackage{color}
\usepackage{hyperref}

\hypersetup{
    colorlinks=true,
    linkcolor=red,
    citecolor=blue,
}

\def\fun#1#2{\lower3.6pt\vbox{\baselineskip0pt\lineskip.9pt
  \ialign{$\mathsurround=0pt#1\hfil##\hfil$\crcr#2\crcr\sim\crcr}}}
\def\simgt{\mathrel{\lower0.6ex\hbox{$\buildrel {\textstyle >}
 \over {\scriptstyle \sim}$}}}
\def\simlt{\mathrel{\lower0.6ex\hbox{$\buildrel {\textstyle <}
 \over {\scriptstyle \sim}$}}}

\def\bea{\begin{eqnarray}}
\def\eea{\end{eqnarray}}
\def\be{\begin{equation}}
\def\ee{\end{equation}}

\input epsf

\newcommand{\mpcoh}{\,h^{-1}\,{\rm Mpc}}

\def\be{\begin{equation}}
\def\ee{\end{equation}}
\def\ba{\begin{eqnarray}}
\def\ea{\end{eqnarray}}

\begin{document}

\preprint{}

\title{The Vainshtein Mechanism in the Cosmic Web}

\author{Bridget Falck$^{(1)}$, Kazuya Koyama$^{(1)}$, Gong-bo Zhao$^{(2,1)}$ and Baojiu Li$^{(3)}$}

\bigskip

\vskip1cm

\affiliation{
\vskip0.3cm
$^{(1)}$Institute of Cosmology \& Gravitation, University of Portsmouth, Dennis Sciama Building, Portsmouth, PO1 3FX, United Kingdom \\
$^{(2)}$
National Astronomy Observatories, Chinese Academy of Science, Beijing, 100012, P. R. China
\\
$^{(3)}$
Institute for Computational Cosmology, Physics Department, University of Durham, South Road, Durham, DH1 3LE, United Kingdom
}


\begin{abstract}
We investigate the dependence of the Vainshtein screening mechanism on the cosmic web morphology of both dark matter particles and halos as determined by ORIGAMI. Unlike chameleon and symmetron screening, which come into effect in regions of high density, Vainshtein screening instead depends on the dimensionality of the system, and screened bodies can still feel external fields. ORIGAMI is well-suited to this problem because it defines morphologies according to the dimensionality of the collapsing structure and does not depend on a smoothing scale or density threshold parameter. We find that halo particles are screened while filament, wall, and void particles are unscreened, and this is independent of the particle density. However, after separating halos according to their large scale cosmic web environment, we find no difference in the screening properties of halos in filaments versus halos in clusters. 
We find that the fifth force enhancement of dark matter particles in halos is greatest well outside the virial radius. We confirm the theoretical expectation that even if the internal field is suppressed by the Vainshtein mechanism, the object still feels the fifth force generated by the external fields, by measuring peculiar velocities and velocity dispersions of halos. Finally, we investigate the morphology and gravity model dependence of halo spins, concentrations, and shapes.
\end{abstract}

\maketitle

\section{Introduction}

The late time acceleration of the Universe is one of the most intriguing mysteries of modern cosmology. The acceleration could simply be caused by the cosmological constant, or alternatively, modifications to General Relativity (GR) on cosmological scales may account for the acceleration. The main problem of modified gravity models is that new degrees of freedom are introduced when GR is modified. These new degrees of freedom are generally scalar degrees of freedom, and they mediate the so-called ``fifth force,'' which is strongly constrained by the solar system experiments. There has been interesting progress in developing ``screening'' mechanisms to suppress scalar interactions on solar system scales. Screening mechanisms invoke non-linearity in the field equations and change the behaviour of the fifth force in high density regions. For example, the chameleon mechanism makes the mass of the field large in high density environments \cite{Khoury:2003rn} whereas the symmetron mechanism changes its coupling to matter \cite{Hinterbichler:2010es}.

One of the oldest ideas to suppress the fifth force is the Vainshtein screening mechanism \cite{Vainshtein}, originally discovered in the context of massive gravity. Massive gravitons have five polarisations instead of the two in GR, and the helicity-0 mode mediates the fifth force. In a linear approximation, this helicity-0 mode does not decouple in the massless limit, leading to the so-called van Dam-Veltman-Zakharov discontinuity \cite{vanDam, Zakharov}. This problem can be solved by the Vainshtein mechanism. If the graviton mass is small, the derivative self-interactions of the helicity-0 mode become important at much larger distances compared with the Schwarzschild radius of a source, and they suppress the coupling of the helicity-0 mode to matter. The key feature of this mechanism is that derivative self-interactions of the scalars are responsible for hiding the fifth force. We do not require any particular form of the scalar potential and any couplings of the scalar to matter, unlike the chameleon and symmetron mechanisms. 

In a simple extension of the Fierz-Pauli massive gravity \cite{Fierz}, the non-linear interactions involve higher derivatives, and this leads to the Boulware-Deser ghost \cite{Boulware}. The non-linear interactions that give an equation of motion containing only up to second derivatives have been constructed imposing a Galilean symmetry \cite{Nicolis:2008in}, and this has led to the discovery of a Boulware-Deser \cite{Boulware} ghost-free massive gravity theory \cite{drgt}. The Vainshtein mechanism occurs not only in massive gravity \cite{Koyama:prd, Koyama:Vainshtein} but also in galileon cosmology \cite{Chow:2009fm, Silva:2009km} and braneworld models. Indeed it is in the braneworld model of  Dvali, Gabadadze, and Porrati (DGP) \cite{Dvali:2000hr} that it has been best studied. See Ref.~\cite{Babichev:2013usa} for a review and references therein. 

Interestingly, these screening mechanisms are distinguished by how screened bodies fall in external fields \cite{Hui:2009kc}.  As a consequence of universal coupling, all un-self-screened test bodies fall in the same way and obey a microscopic equivalence principle. In the chameleon and symmetron models, screened bodies do not respond to external fields while in the Vainshtein mechanism they do, as long as those fields have wavelengths long compared to the Vainshtein radius \cite{Hui:2012jb}. These differences arise because of the non-superimposability of field solutions. Also, the peculiar structure of the non-linear interactions in the Vainshtein mechanism implies that the screening depends on the dimensionality of the system. For example, the Vainshtein mechanism does not work at all in one-dimensional systems \cite{Brax:2011sv}.

The potential dimensionality dependence of the Vainshtein screening mechanism motivates us to look for its signatures in the cosmic web of large scale structure. On very large scales, the distribution of matter in the present-day Universe forms into complex, interconnected, hierarchical structures composed of voids, walls or sheets, filaments, and halos or knots. While voids are underdense regions and halos are density peaks, it is now understood that density alone is not the salient feature that distinguishes these structures. Rather, it is the dynamics of the nonlinear gravitational collapse that fundamentally distinguishes between expanding voids, walls collapsing along one dimension, filaments collapsing along two, and halos collapsing along three orthogonal dimensions \cite{Zeldovich:1969sb, Klypin, Bond:1995yt, AragonCalvo:2007mk, Hahn:2006mk, Falck, Hoffman:2012ft}. While the chameleon mechanism has been found to depend on the density of the halo environment 
\cite{Zhao:2011cu}, the dimensionality dependence of the Vainshtein mechanism suggests it may depend instead on the morphology of the cosmic web.

In this paper, we study how the Vainshtein mechanism operates to hide the fifth force in the cosmic web of large scale structure using cosmological $N$-body simulations. The most general effective theory in four dimensions without involving higher derivative operators was derived in Ref.~\cite{Koyama:2013paa}. Although there are many parameters in this effective theory, the stability condition around the spherically symmetric solutions significantly restricts the available parameter space. For this study, we focus on the simplest non-linear operator, the cubic Galileon term, which appears in the DGP braneworld and galileon models~\cite{Barreira:2013eea}. We consider the normal branch DGP model where the background expansion history is exactly the same as LCDM to disentangle the effects of different cosmological backgrounds and those of the Vainshtein mechanism.

The paper is organised as follows. In section II, we introduce a normal branch DGP model and describe the evolution of the background cosmology and quasi-static perturbations. We provide spherically symmetric solutions for a dark matter halo described by the NFW profile~\cite{Navarro:1996gj}. Then we describe our $N$-body simulations in detail. In section III, we study the fifth forces acting on dark matter particles. We describe the ORIGAMI method of Ref.~\cite{Falck} to identify the morphology of dark matter particles and study the dependence of the Vainshtein mechanism on the morphology. In section IV, we study dark matter halos. We first identify dark matter halos using the ORIGAMI code, determine their cosmic web environment, and study the screening of dark matter halos. We then study velocity dispersions and peculiar velocities of halos to investigate how screened bodies respond to external fields. Finally, we study the morphological environment and screening dependence of various halo properties by cross-matching ORIGAMI halos with halos found using the AHF code~\cite{Gill:2004km,Knollmann:2009pb}. Section V is devoted to conclusions.

\section{Model and Simulations}

\subsection{Model}

In order to disentangle the effects of different cosmological backgrounds and those of the Vainshtein mechanism, we consider the normal branch DGP (nDGP) braneworld model that has exactly the same expansion history as the LCDM model. Note that our aim is not to test this specific model but to identify characteristic features of the Vainshtein mechanism. We expect that our findings can be applied to other models that accommodate the Vainshtein mechanism.

The Friedman equation in nDGP is given by
\begin{equation}
H^2 = -\frac{H}{r_c} + \frac{8 \pi G \rho_m}{3} +\frac{8 \pi G \rho_{\rm DE}}{3} ,
\end{equation}
where $r_c$ is the cross-over scale at which gravity transitions from being 5D on large scales to 4D on small scales. We tune the equation of state of dark energy so that the background expansion history becomes the same as LCDM \cite{Schmidt:2009sv}. Thus deviations from LCDM are solely caused by the modified growth due to the extra scalar degree of freedom in the model. We assume dark energy does not cluster on sub-horizon scales. The background expansion history is given by
\begin{equation}
H^2 = \frac{8 \pi G \rho_m}{3} +  \frac{8 \pi G \rho_{\Lambda}}{3}.
\end{equation}

Under the quasi-static perturbations, the Poisson equation and the equation for the scalar field are given by \cite{Koyama:2007ih}
\begin{equation}
\nabla^2 \Psi = \nabla^2 \Psi_N + \frac{1}{2} \nabla^2 \varphi,
\end{equation}
and
\begin{equation}
\nabla^2 \varphi + \frac{r_c^2}{3 \beta(a) a^2}
[(\nabla^2 \varphi)^2 -(\nabla_i \nabla_j \varphi)(\nabla^i \nabla^j \varphi)]
= \frac{8 \pi G a^2}{3 \beta(a)} \rho \delta,
\label{eq:phievo}
\end{equation}
where $\Psi$ is the Newtonian potential, and we define the Newtonian potential in GR as
\begin{equation}
\nabla^2 \Psi_N = 4 \pi G a^2 \rho \delta.
\end{equation}
The function $\beta(a)$ is given by
\begin{equation}
\beta(a) = 1 +  2 H r_c \left(1+ \frac{\dot{H}}{3 H^2} \right).
\label{eq:beta}
\end{equation}
If we linearise the equations, the Poisson equation is given by
\begin{equation}
\nabla^2 \Psi = 4 \pi G a^2 \left(1+ \frac{1}{3 \beta(a)} \right) \rho \delta.
\label{linear}
\end{equation}
Note that $\beta$ is always positive, so the growth of structure formation is enhanced in this model.

This model has one extra parameter, $r_c$, in addition to usual cosmological parameters in the LCDM model. If $r_c$ becomes  larger, the enhancement of gravity is weaker and also the Vainshtein mechanism operates more efficiently as the amplitude of the non-linear terms are larger. Thus in this limit we recover LCDM.

\subsection{Spherically Symmetric Solution}
\label{sec:spherical}
In order to understand how the Vainshtein mechanism operates, it is useful to consider spherically symmetric solutions \cite{Schmidt:2009yj, Schmidt:2010jr}.
For a spherically symmetric object, the scalar field equation (\ref{eq:phievo}) reduces to
%
\begin{equation}
  3\left(\frac{d^2\varphi}{dr^2} + \frac{2}{r}\frac{d\varphi}{dr}\right)
  + r_c^2\left[\frac{2}{r^2}\left(\frac{d\varphi}{dr}\right)^2
  + \frac{4}{r}\frac{d^2\varphi}{dr^2}\frac{d\varphi}{dr}\right]
  = 8 \pi G \delta\rho.
\label{eq:phisph}
\end{equation}
Eq.~(\ref{eq:phisph}) can be integrated by multiplying both
sides by $r^2 dr$, resulting in
%
\begin{equation}
 3 r^2\frac{d\varphi}{dr}
     + 2 r r_c^2\left(\frac{d\varphi}{dr}\right)^2
=2 G m(r),
\end{equation}
where we define the enclosed mass fluctuations $m(r)$ as
%
\begin{equation}
m(r)= 4\pi  \int_0^r\!\delta\rho(r')r'^2\,dr'.
\end{equation}
%
As this is a quadratic equation for $d\varphi/dr$ we immediately obtain 
%
\begin{equation}
  \frac{d\varphi}{dr} = \frac{G m(r)}{r^2} \frac{4}{3 \beta} g\left(\frac{r}{r_*} \right), \quad
g(x) = x^3 \left( \sqrt{1+x^{-3}} -1 \right),
\label{eqn:Birkhoff}
\end{equation}
%
and
\begin{equation}
r_*= \left(\frac{16 G m(r) r_c^2}{9 \beta^2}\right)^{1/3}.
\label{r*}
\end{equation}
%
The Newtonian potential in GR is calculated as
%
\begin{equation}
\frac{d \Psi_N}{d r} =  \frac{G m(r)}{r^2}.
\end{equation}
%
The radius $r_*$ outside the matter distribution $m(r)=$ const. is called the Vainshtein radius. The Vainshtein radius depends on the gravitational length of the object, $r_g = 2 G m$, and the cross-over scale, $r_c$. Inside the Vainshtein radius, the scalar force is suppressed compared to the Newtonian potential and we recover the Newtonian solution. For a larger $r_c$, the Vainshtein radius is larger, thus the region in which the fifth force is suppressed becomes larger and we recover GR. This recovery of GR is achieved by the non-linearity of the scalar field.
Note that Birkhoff's theorem does not hold in the DGP model in general because the gravitational force on a test particle due to a spherical mass shell depends on the mass distribution~\cite{Dai:2008zza}. However, for a solution that satisfies the regularity condition in the fifth dimension, Birkhoff's theorem holds effectively, since the gravity at a given radius $r$ is determined by the interior mass for a spherically symmetric mass distribution, as seen from the solution in Eq.~(\ref{eqn:Birkhoff}).

We define the ratio between the fifth force and the Newtonian force as
\begin{equation}
\Delta_M = \frac{1}{2} \frac{d \varphi/ dr} {d \Psi_N/dr}.
\label{eqn:deltam}
\end{equation}
For a spherically symmetric object, this is given by
\begin{equation}
\Delta_M = \frac{2}{3 \beta} g\left(\frac{r}{r_*} \right).
\end{equation}
For large $r$, $g(x)  \to 1/2$ thus we recover the linear prediction $\Delta_M = 1/3 \beta$. On the other hand, for small $r$, $g(x) \to 0$ and the fifth force is suppressed.

We are interested in a dark matter halo modeled by the NFW profile \cite{Navarro:1996gj} with a mass $M_{\Delta}$. The mass is defined as the mass contained within the radius $r=r_{\Delta}$; the density at $r_{\Delta}$ is $\rho_{\rm crit} \Delta$, where $\rho_{\rm crit}$ is the critical density of the Universe. The NFW profile is given by
\begin{equation}
\rho(r) = 4 \rho_s f \left(\frac{r}{r_s} \right), \quad
f(y) = \frac{1}{y (1+y)^2} ,
\end{equation}
where $\rho_s =\rho(r_s)$ is fixed so that the mass within $r_{\Delta}$ is $M_{\Delta}$. The remaining parameter $r_s$ is more conveniently parameterised by the concentration $c_{\Delta} = r_{\Delta}/r_s$. By integrating this density profile, we obtain $m(r)$ as
\begin{equation}
m(r) = M_{\Delta} \frac{F (c_{\Delta} r/r_{\Delta})}{F(c_{\Delta})}, \quad
F(y)= -\frac{y}{1+y} + \ln(1+y),
\label{NFWm}
\end{equation}
where we neglected a term suppressed by $\Delta$. Substituting this expression into the definition of $r_*$, we can calculate $\Delta_M$ in a halo with the mass $M_{\Delta}$ and the concentration $c_{\Delta}$. These theoretical predictions will be tested in Section~\ref{sec:origamihalos}.

\subsection{Simulations}

The simulations for models with the Vainshtein mechanism were developed in Ref.~\cite{Li:2013nua}. This code, dubbed ECOSMOG-V, is a variant of the ECOSMOG code~\cite{Li:2011vk}, which has been used to simulate the DGP and Galileon models~\cite{Barreira:2013eea, Li:2013tda}. The main features of the code are summarised as follows:
\begin{enumerate}
\item It solves the scalar field on a mesh using the Newton-Gau{\ss}-Seidal nonlinear relaxation, which has good convergence properties. The density field on the mesh is obtained by assigning particles following the triangular-shaped-cloud (TSC) scheme, and there is no need to pre-smooth it to achieve convergence;
\item The mesh can be adaptively refined in high-density regions to achieve higher resolution and accuracy there, without affecting the overall performance. There is no need to smooth the density field on the refinements either; and
\item It is efficiently parallelised using {\sc mpi}, which makes the simulations fast.
\end{enumerate}
We refer to Ref.~\cite{Li:2013nua} for details. 

The baseline cosmology for the nDGP simulations was chosen to be the best-fit Planck cosmological parameters (Table 9 in \cite{Planck2013} for the joint dataset of Planck + WP + highL + BAO): $\Omega_b h^2=0.022161,~\Omega_c h^2=0.11889,~\Omega_K =0,~h=0.6777, n_s=0.9611,~A_s=2.21381\times10^{-9}$. We simulate three nDGP models: nDGP1 ($H_0 r_c = 0.57$), nDGP2 ($H_0 r_c= 1.1987$), and nDGP3 ($H_0 r_c = 5.65$). The values of $r_c$ are tuned so that these models have correspondence with the $f(R)$ model of Ref.~\cite{Hu:2007nk} (Hu-Sawicki) in terms of $\sigma_8$. Specifically, nDGP1, 2, and 3 have the same $\sigma_8$ as the Hu-Sawicki model with $|f_{R0}|=10^{-4}$, $10^{-5}$, and $10^{-6}$, respectively. This is for the purpose of comparing the Vainshtein with the chameleon screening mechanism, which will be the subject of future work.

We run $N$-body simulations using $256^3$ particles in a $L=64\mpcoh$ box from the initial redshift $z=49$ to $z=0$. The initial conditions are generated using MPGrafic~\cite{mpgrafic}, which is a parallel version of the Grafic code~\cite{grafic}. Three realisations are simulated for the same model to reduce the sample variance, and we measure the matter power spectrum $P(k)$ from the simulation using POWMES~\cite{powmes}, which is a numeric tool for $P(k)$ measurement to a high precision.

\begin{figure}[h]
  \centering{
  \includegraphics[width=10cm]{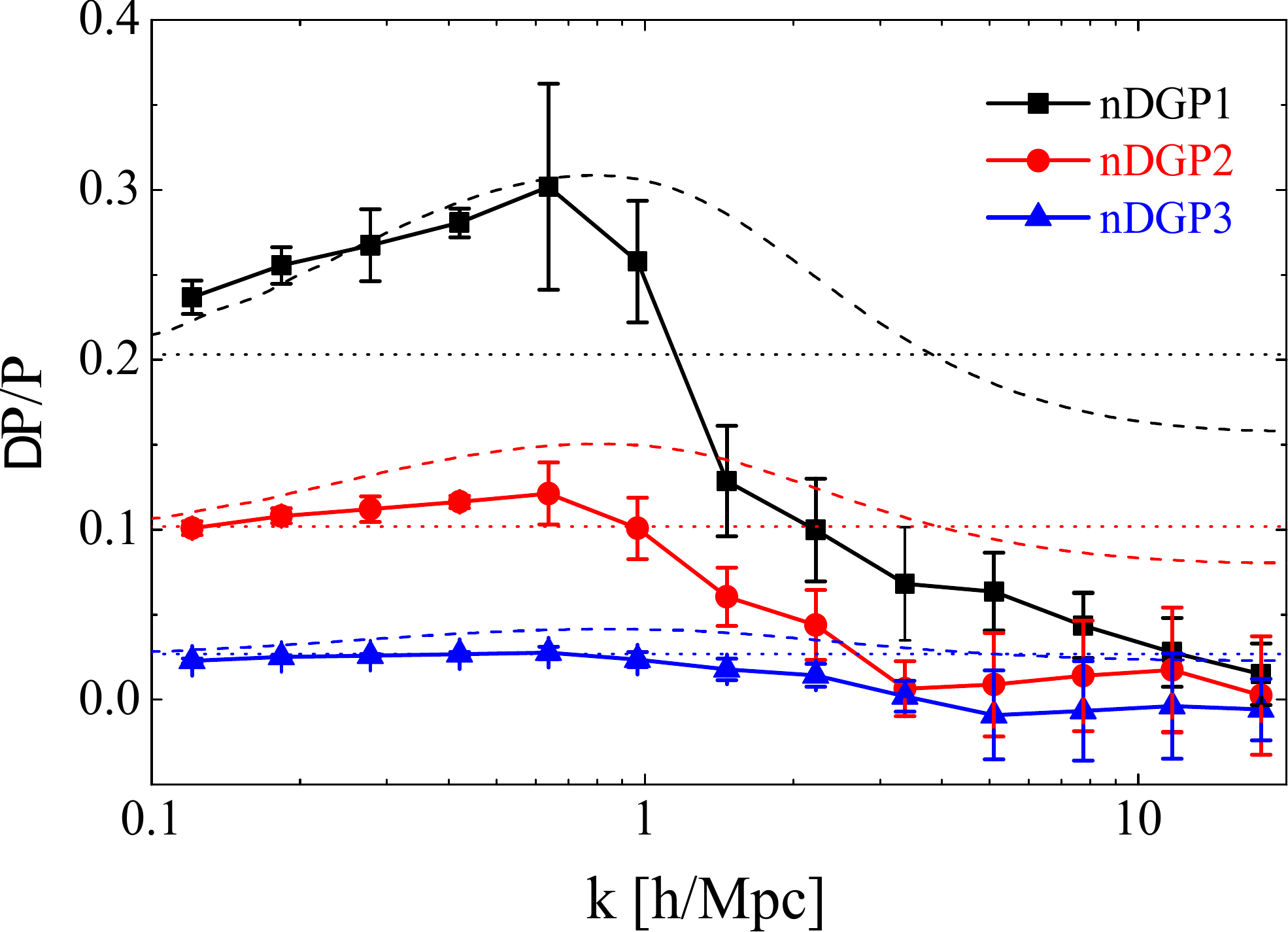}
  }
  \caption{The power spectra of the three nDGP models shown as the fractional difference with the LCDM simulations at $z=0$ (solid lines). Error bars come from the variance of the three simulation realisations. The enhancements in the power spectra on large scales agree well with the linear theory predictions, the Halofit model (dashed lines) captures the transition from linear to non-linear scales, and on small scales the Vainshtein mechanism suppresses the enhancement of power, approaching zero at large $k$.
  }
  \label{fig:power}
\end{figure}

Fig.~\ref{fig:power} shows the fractional difference of the power spectrum in nDGP models compared to LCDM, measured from the simulations at $z=0$. The dotted line shows the linear prediction while the dashed line is the prediction obtained by using the Halofit model~\cite{Smith:2002dz} that gives a mapping from the linear power spectrum to the non-linear one. On large scales, the power spectrum enhancements agree well with the linear theory predictions. On quasi-linear scales, the Halofit model captures the transition from linear to non-linear scales. On small scales, the Vainshtein mechanism suppresses the enhancement, and the fractional difference eventually approaches zero at large $k$. Halofit is developed in GR models, thus it does not include the suppression of the enhancement due to the Vainshtein mechanism and significantly over-estimates the power on small scales in nDGP models.

The excellent agreement with linear theory on large scales is one of the key features of the Vainshtein mechanism. In other screening mechanisms, such as the chameleon mechanism, once the fifth force is suppressed inside dark matter halos, these screened halos no longer feel the fifth force. In these models, the scalar field is massive, the fifth force is suppressed beyond the Compton wavelength of the scalar field, and the enhancement of the power spectrum is suppressed on large scales. In addition, the chameleon mechanism further suppresses the enhancement of the power spectrum even on linear scales due to the screening of dark matter halos. On the other hand, in models with the Vainshtein mechanism, even if the fifth force is suppressed inside halos, these halos still feel external scalar fields as long as those fields have wavelengths long compared to the Vainshtein radius. Thus the linear theory works well on large scales. We will confirm this picture later by studying the velocities of dark matter particles inside halos (see Section~\ref{sec:velocity}).

\section{Dark Matter Particles}

In this section, we study how the Vainshtein mechanism operates for dark matter particles in simulations. We calculate the fifth force at the positions of the dark matter particles, and we also compute the gravitational force at the particle positions. If there are no non-linear interactions of the scalar field, then there is a linear relation between the fifth force and the gravitational force as predicted by the linear theory, i.e. $\Delta_M=1/3 \beta$. Deviations from this linear relation indicate that the Vainshtein mechanism is working.

The Vainshtein mechanism is based on the non-linear terms in the equation of motion $(\nabla^2 \varphi)^2 -(\nabla_i \nabla_j \varphi)(\nabla^i \nabla^j \varphi)$. This non-liner term has an interesting property. For the spherically symmetric case with a top-hat density source, we have the relation
\begin{equation}
(\nabla^2 \varphi)^2 -(\nabla_i \nabla_j \varphi)(\nabla^i \nabla^j \varphi) =
\frac{2}{3} (\nabla^2 \varphi)^2.
\end{equation}
On the other hand, if we consider a one-dimensional system, the non-linear term identically vanishes. This implies that the Vainshtein mechanism depends on the cosmic web morphology of the matter distribution.

\subsection{ORIGAMI}

We utilise the ORIGAMI code\footnote{\url{http://icg.port.ac.uk/~falckb}} developed in Ref.~\cite{Falck} to identify the morphology of dark matter particles. The large scale structure can be separated into four distinct morphologies, characterized primarily by their dimensions: halos are point-like clumps, filaments are one-dimensional strands, walls are two-dimensional sheets, and voids are roughly spherical underdense regions. This cosmic web of large scale structure forms through the gravitational collapse of dark matter in the Universe and is well described on large scales with the Zel'dovich approximation~\cite{Zeldovich:1969sb, Bond:1995yt}. In the Zel'dovich picture of structure formation, the three-dimensional nature of collapse is captured by the three eigenvalues of the tidal tensor (the Hessian of the gravitational potential) or the shear tensor (the Hessian of the velocity potential), whose relative values determine the topology of the evolving density field. Thus, though density correlates with morphology, with halos forming in high-density regions of the density field and voids in the underdense regions, the structure of the cosmic web is governed primarily by dynamics.

When dark matter particles move toward an over-density to form structures, at some point they cross paths with other particles at a caustic. ORIGAMI uses the information of the initial locations of dark matter particles in simulations to determine whether this crossing has occurred, which denotes the formation of caustics within which the velocity field is multi-valued. If crossings have occurred along three orthogonal axes, the particle is tagged as a halo particle with a morphology index $M=3$. Similarly, filament particles have crossed along two orthogonal axes ($M=2$), walls along one ($M=1$), and void particles have not crossed any caustics ($M=0$). Thus each particle in the simulation is identified by its cosmic web morphology in a dynamical, scale-independent way, without relying on the density field.

Once all particles have been given a morphology classification, we also compute the Voronoi Tessellation Field Estimate (VTFE) to determine a density for each particle~\cite{Schaap:2000se, vandeWeygaert:2007ze}. A Voronoi tessellation partitions space into cells, such that all points inside a particle's Voronoi cell are closer to that particle than to any other. The Delaunay tessellation is the dual of the Voronoi tessellation and divides a three-dimensional volume into a set of tetrahedra that connect particles, such that the particles in two adjacent Voronoi cells are connected in the Delaunay tessellation. The VTFE density at each particle is given by $\delta_{\rm VTFE} =\bar{V}/V-1$, where $V$ is the particle's Voronoi cell volume and $\bar{V}$ is the average of $V$ among all particles.

\begin{figure}[ht]
  \centering{
  \includegraphics[width=15cm]{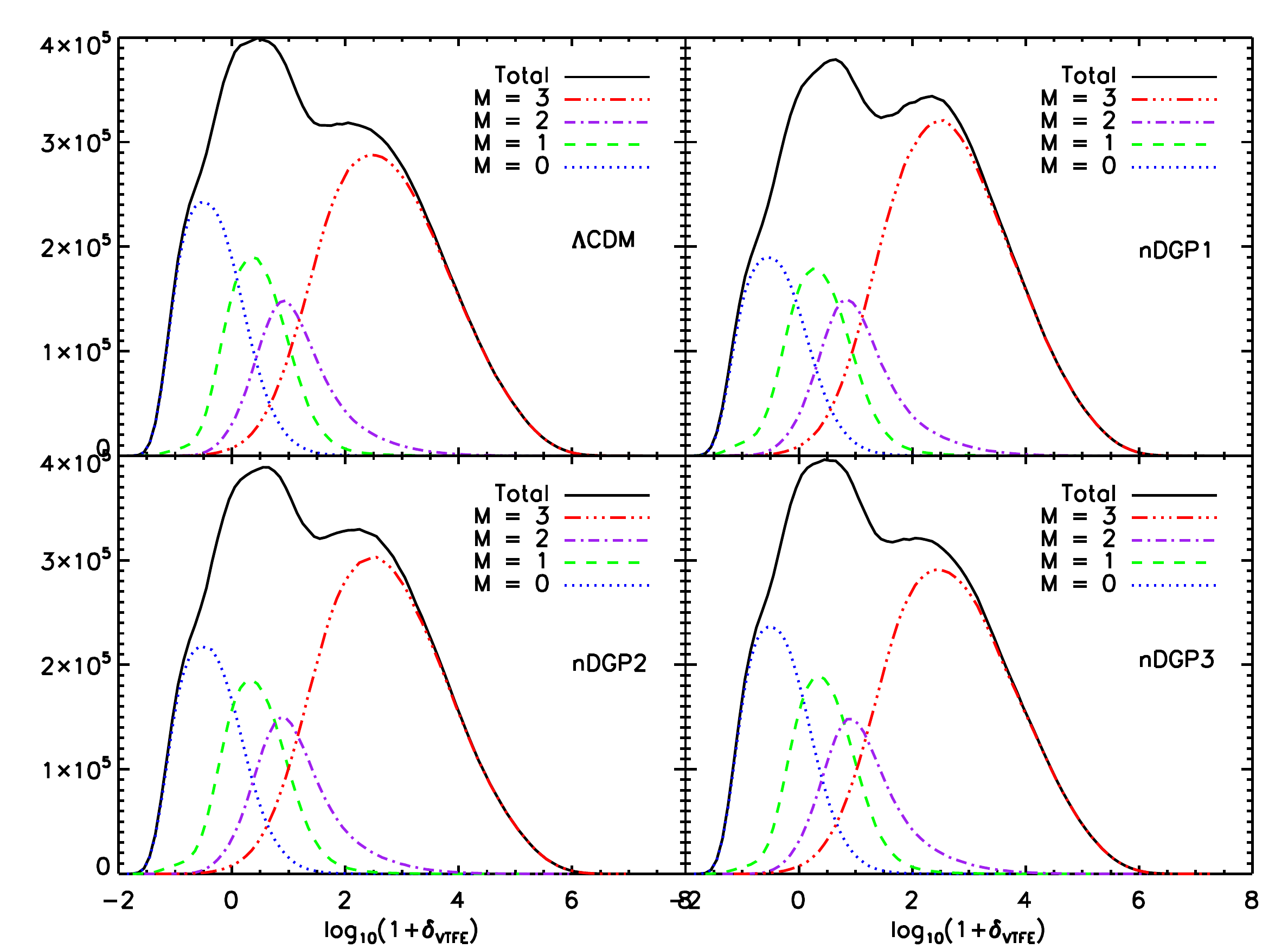}
  }
  \caption{Distribution of particle densities measured using the Voronoi tessellation, separated by ORIGAMI morphology, for one realization of the LCDM simulation and the three nDGP models under study. nDGP1 (top right panel) has the strongest enhancement to gravity, resulting in a larger high-density peak of halo particles, while nDGP3 (bottom right panel) is very similar to LCDM (top left panel). The ORIGAMI morphology index $M$ is 0, 1, 2, and 3 for void, wall, filament, and halo particles respectively.
  }
  \label{fig:vtfemorph}
\end{figure}

Figure~\ref{fig:vtfemorph} shows the VTFE density distribution for ORIGAMI particles separately based on their morphology index $M$, where again $M = 0$, 1, 2, and 3 corresponds to void, wall, filament, and halo respectively. Because the VTFE density distribution counts particles instead of volumes (as would a grid-based density like cloud-in-cell), the distribution function has a characteristic high-density bump that becomes more prominent as the simulation resolution increases (see, e.g., \cite{Falck}). This is a reflection of the fact that as resolution increases, more halos are capable of being resolved (at least if dark matter is cold), which can populate low-density regions inside walls and voids.

Figure~\ref{fig:vtfemorph} shows that the strength of gravity also has an effect on the shape of the VTFE density distribution, such that the model with the most gravity enhancement, nDGP1, has more high-density particles and more halo particles, and similarly fewer low-density and void particles. 
These differences are quantified in Table~\ref{tab:massfrac}, which gives the fractions of void, wall, filament, and halo particles according to gravity model, averaged over the three independent simulation realizations. nDGP1 has fewer void particles and more halo particles, while nDGP3 is very close to LCDM.

\begin{table}[ht]
\centering
\begin{tabular}{c|cccc}
Gravity Model & Void & Wall & Filament & Halo\\
\hline
LCDM &  0.19  &   0.16   &   0.14  &   0.51 \\
nDGP1 &  0.15  &   0.16   &   0.14  &   0.55 \\
nDGP2 &  0.17  &   0.16   &   0.14  &   0.53 \\
nDGP3 &  0.19  &   0.16   &   0.14  &   0.51 \\
\end{tabular}
\caption{Mass fractions of particles with each ORIGAMI morphology, averaged over the three independent simulation realizations, for each gravity model.}
\label{tab:massfrac}
\end{table}

In addition to the mass fractions of the different morphological types, we look at the average VTFE log-density, $\langle\log(1+\delta)\rangle$, of void, wall, filament, and halo particles for the different gravity models, averaged over the three independent simulation realizations, in Table~\ref{tab:avgden}. This shows a small tendency for nDGP models to have particles with lower VTFE densities, for all four morphologies, but note that the distributions of VTFE densities for each morphology are much wider than these small changes in the mean (see Figure~\ref{fig:vtfemorph}).

\begin{table}[ht]
\centering
\begin{tabular}{c|cccc}
Gravity Model & Void & Wall & Filament & Halo\\
\hline
LCDM &  -0.34  &   0.40   &   1.12  &   2.73 \\
nDGP1 &  -0.40  &   0.27   &   1.02  &   2.68 \\
nDGP2 &  -0.36  &   0.35   &   1.09  &   2.71 \\
nDGP3 &  -0.34  &   0.39   &   1.11  &   2.72 \\
\end{tabular}
\caption{Average VTFE log-density, $\langle\log(1+\delta)\rangle$, of particles with each ORIGAMI morphology, averaged over each of the three independent simulation realizations, for each gravity model.}
\label{tab:avgden}
\end{table}

\subsection{Density, Morphology, and Fifth Force}

\begin{figure}[ht]
  \centering{
  \includegraphics[width=15cm]{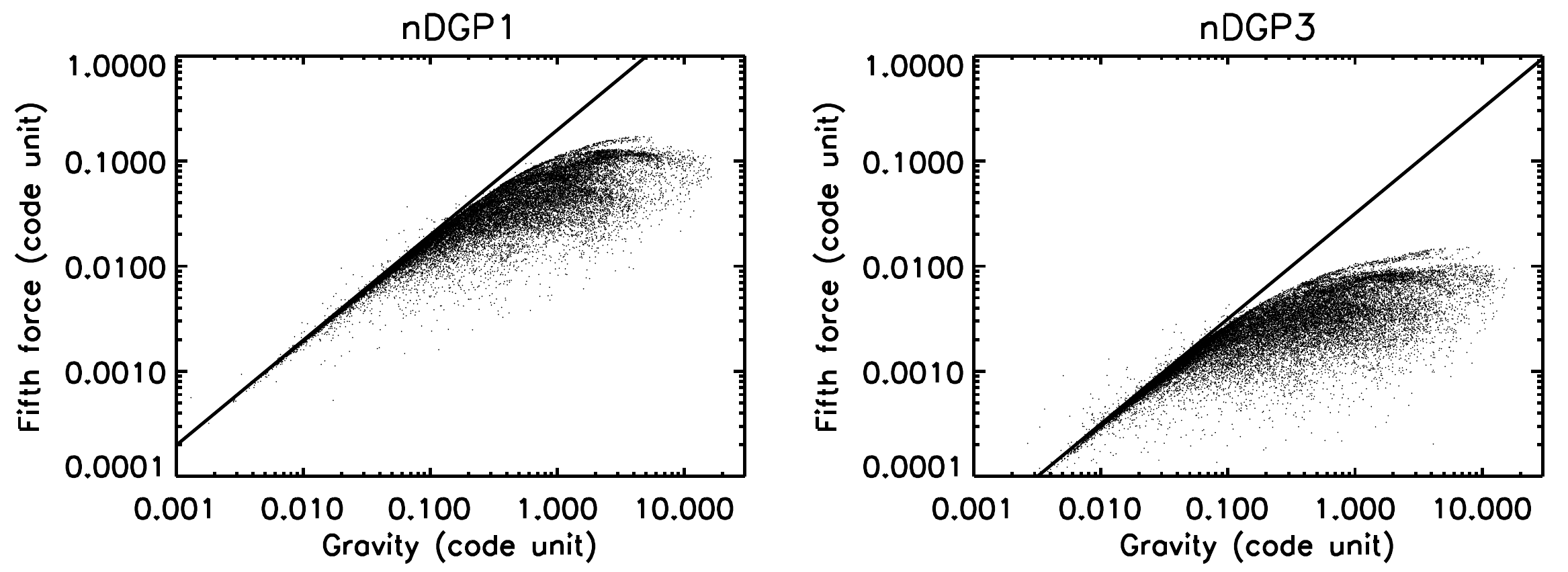}
  }
  \caption{The fifth force vs. gravity for a subset of 50000 dark matter particles in one realization of the nDGP1 (left) and nDGP3 (right) simulations. The unscreened particles fall along the linear prediction, given by the straight line, while the Vainshtein mechanism acts to reduce the fifth force when gravity is strong, causing particles to fall below the linear prediction. 
  }
  \label{fig:fifthforce}
\end{figure}

In this section we study the morphology and density dependence of the fifth force and gravity acting on each particle. Figure~\ref{fig:fifthforce} shows the fifth force vs. gravitational force for a subset of particles in two nDGP models, with the straight line giving the linear theory ratio $\Delta_M=1/3 \beta$. As expected, the fifth force is strongly suppressed when gravity is strong as the Vainshtein mechanism is effective. On the other hand, the fifth force vs. gravity obeys the linear relation when gravity is weak. In nDGP1 (left panel), where the gravity enhancement is stronger, this ratio is higher, and particles experience a higher fifth force on average compared to nDGP3 (right panel).

\begin{figure}[ht]
  \centering{
  \includegraphics[width=15cm]{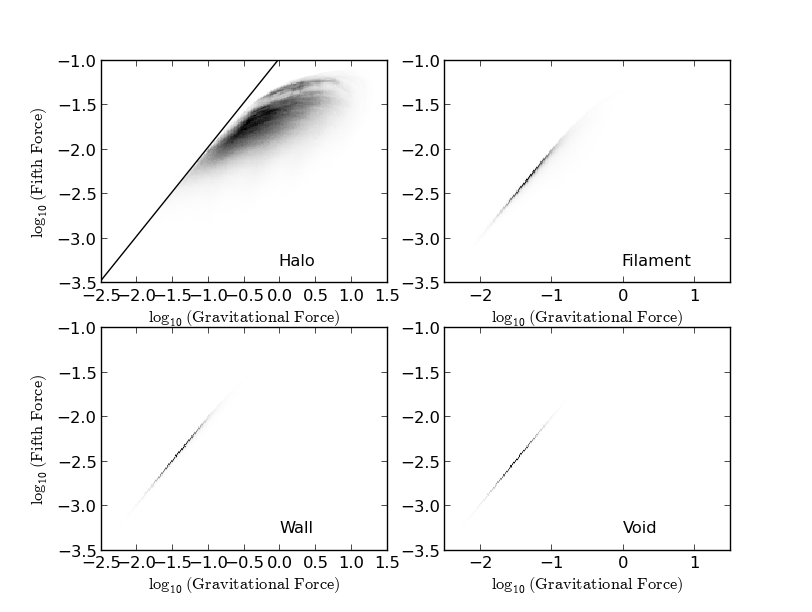}
  }
  \caption{The fifth force vs. gravity for dark matter particles in an nDGP2 simulation split according to ORIGAMI morphology. Halo particles (upper left) are screened while filament (upper right), wall (lower left), and void (lower right) particles fall along the linear theory prediction (given by the straight line in the upper left panel only, for clarity).
  }
  \label{fig:force-morph}
\end{figure}

To determine the effect of density and morphology on the Vainshtein mechanism, we divide particles into four groups according to their VTFE densities and morphologies. Figure~\ref{fig:force-morph} shows a two-dimensional histogram of the fifth force vs. gravitational force for all particles in an nDGP2 simulation, split according to their ORIGAMI morphology. The linear ratio is plotted as a straight line in the halo panel and would fall directly over the particles in the filament, wall, and void panels. It is clear that most of the halo particles are screened, while most of the filament, wall, and void particles are unscreened. (Note that the percent of halo, filament, wall, and void particles in this simulation are 52\%, 14\%, 15\%, and 19\%, respectively.) Thus ORIGAMI, by separating particles according to the dimensionality of their gravitational collapse, is able to distinguish screened from unscreened dark matter particles and identify those particles for which the Vainshtein mechanism is effective.

\begin{figure}[ht]
  \centering{
  \includegraphics[width=15cm]{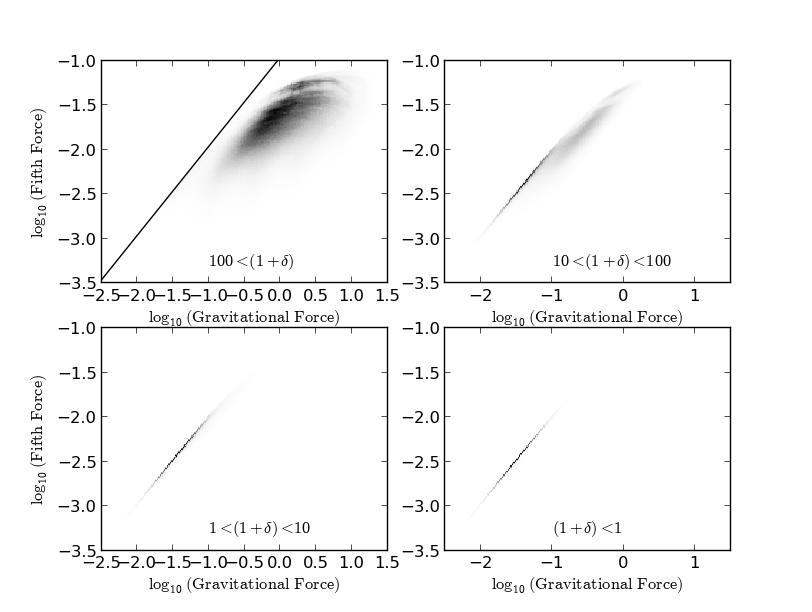}
  }
  \caption{Same as Figure~\ref{fig:force-morph}, but with particles in each panel split according to VTFE density instead of morphology. The highest density bin (upper left) contains only screened particles while the lowest density bins (bottom left and right) contain mostly unscreened particles, and the moderately high density bin (upper right) contains a mix of screened and unscreened particles. The particles in this mixed bin are separated according to morphology in Figure~\ref{fig:force-fixden}.
  }
  \label{fig:force-den}
\end{figure}

Now we look at the fifth force vs. gravitational force for particles split according to their VTFE density. Figure~\ref{fig:force-den} is the same as Figure~\ref{fig:force-morph}, but the four panels instead show particles in four different bins of $\log(1+\delta_{\rm VTFE})$. Again, the linear ratio is plotted only for the upper left panel to avoid confusion. As opposed to splitting by morphology, there is no clean separation of screened and unscreened particles. Though the highest-density bin contains only screened particles and the lowest-density bin contains only unscreened particles, the moderate density bins (especially the upper right panel) contain a mix of screened and unscreened particles, which is not seen in Figure~\ref{fig:force-morph} when splitting by morphology.

\begin{figure}[ht]
  \centering{
  \includegraphics[width=15cm]{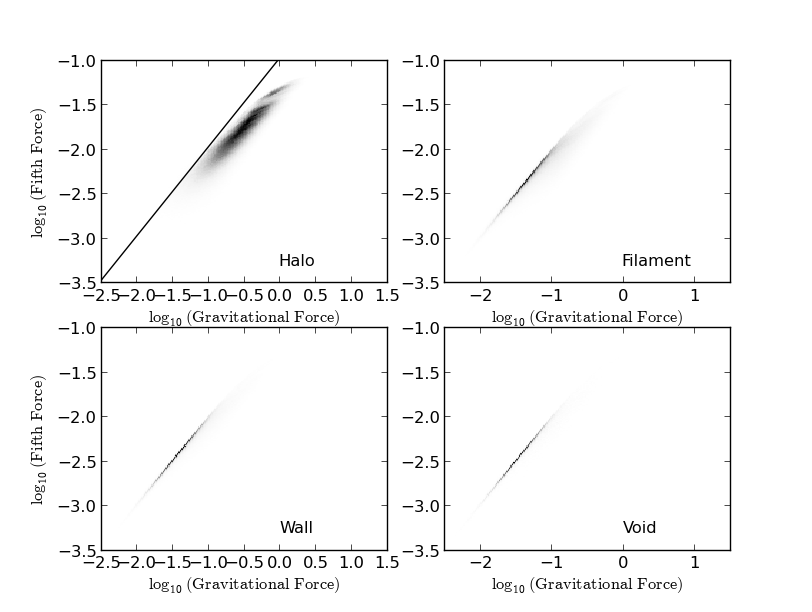}
  }
  \caption{Same as Figure~\ref{fig:force-morph}, but for only those particles with VTFE densities in the range $1 < \log(1+\delta_{\rm VTFE}) < 2$ (upper right panel of Figure~\ref{fig:force-den}). Even at constant density, ORIGAMI morphology is able to distinguish screened halo particles from unscreened dark matter particles.
  }
  \label{fig:force-fixden}
\end{figure}

As shown in Figure~\ref{fig:vtfemorph}, the ORIGAMI morphology index correlates with particle density, with halos having the highest densities and voids the lowest, though there is much overlap. To separate the effects of density and morphology, we look at the morphology dependence of the fifth force for a given density. In Figure~\ref{fig:force-fixden} we show the fifth force vs. gravity of dark matter particles split by ORIGAMI morphology for all particles with VTFE densities in the range $1 < \log(1+\delta_{\rm VTFE}) < 2$ (i.e. the upper right panel of Figure~\ref{fig:force-den}). Again, the ORIGAMI morphology is able to distinguish screened particles, for which the Vainshtein mechanism is effective, from unscreened filament, wall, and void particles.

\section{Dark Matter Halos}

We have shown that identifying the morphology of dark matter particles in simulations according to the dynamics of their collapse can distinguish between screened and unscreened particles under the Vainshtein mechanism. In order to find observable signatures of this effect, however, we have to move from dark matter particles to the halos in which galaxies reside. The halos themselves have different morphologies; they can be in overdense or underdense environments, and they can inhabit different types of structures in the cosmic web. Halos in simulations, and galaxies in observations, are found in filaments, walls, and voids, and obviously there are many in clusters and superclusters 
\cite{AragonCalvo:2007mk, Hahn:2006mk, Tempel:2013wha, Alpaslan:2014ura}. In this section, we investigate the morphological dependence of the Vainshtein mechanism for dark matter halos.

\subsection{ORIGAMI Halos}
\label{sec:origamihalos}
As described in~\cite{Falck}, ORIGAMI halo particles are grouped into halos by associating halo particles that are connected on the Delaunay tessellation, which gives a natural set of nearest neighbors to every particle. ORIGAMI halos extend further than spherical overdensity or friends-of-friends halos that use a density or linking-length parameter to halt their growth; this means that ORIGAMI halos contain particles participating in the collapse that are well outside the halo's virial radius, but this also means that the outer regions of nearby halos can be connected on the tessellation. To prevent over-connected or dumbbell-shaped halos, we first identify halo cores in which each particle is above a density threshold, and then grow halos out from the cores by iteratively adding connected halo particles on the tessellation. For this paper, we use a core VTFE density threshold of $\rho/\bar{\rho}=150$; see~\cite{Falck} for more details on the effect of the core density threshold. Note that any subhalos will be counted as part of the main halo, and we require that halos have a minimum of 20 particles.

We then determine the morphological environment of ORIGAMI halos, i.e. whether they are in a cluster, filament, wall, or void. We first calculate the number of halo neighbors of each halo, again using the Delaunay tessellation to identify neighbors of halo particles. If a halo is connected to three or more different halos on the tessellation, we say that it is in a cluster. If not, we calculate the fraction of halo, filament, wall, and void particles that the halo is connected to and assign the halo to the morphology with the highest fraction of connected particles.

\begin{figure}[t]
  \centering{
  \includegraphics[width=15cm]{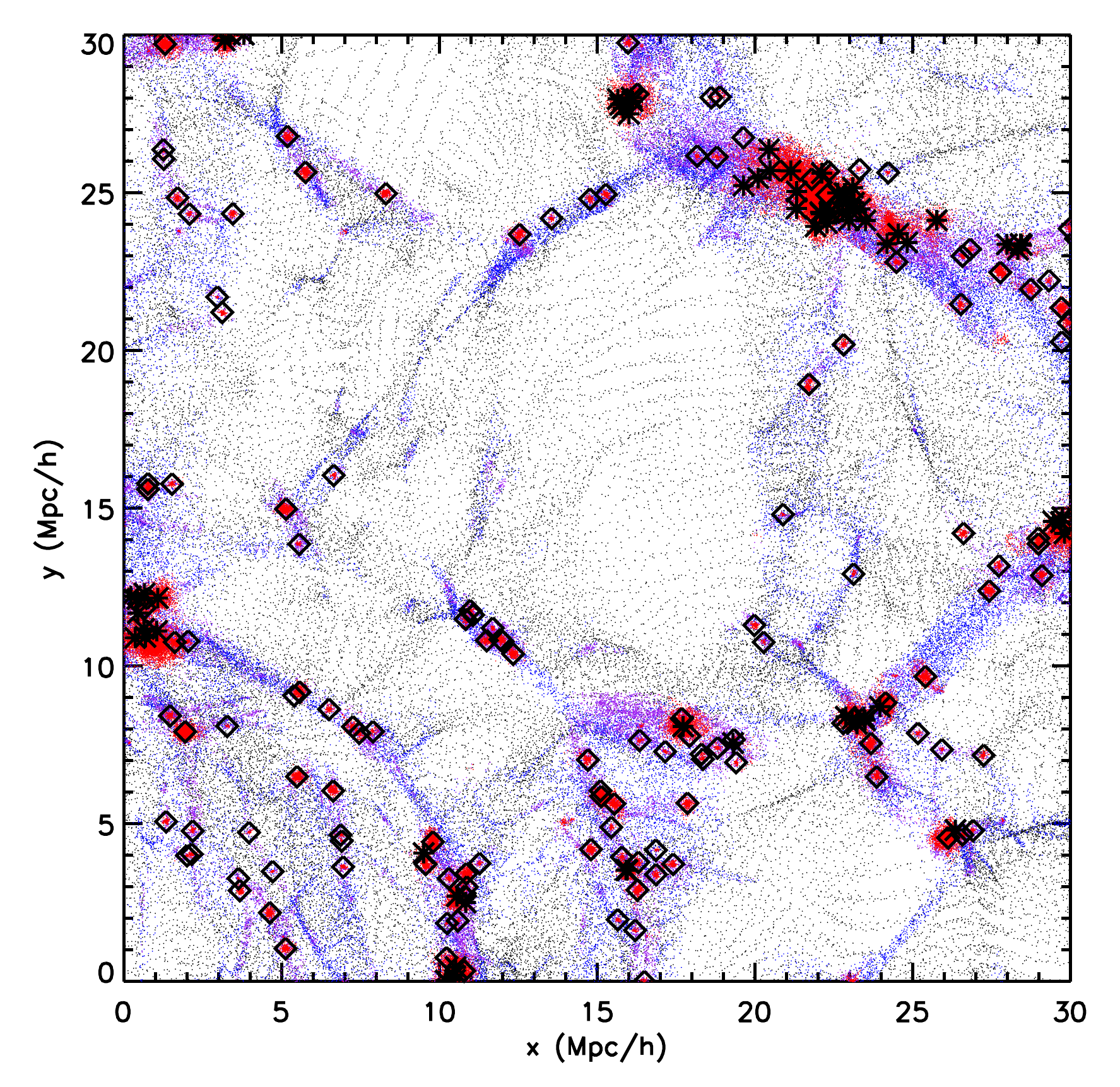}
  }
  \caption{Particles in a 5$\mpcoh$-thick slice of a nDGP2 simulation, showing ORIGAMI halo positions as either stars (cluster halos) or diamonds (filament halos). Void particles are plotted in black, wall in blue, filament in purple, and halo in red (in that order). While the most massive halos are in clusters, most of the halos are in filaments.
  }
  \label{fig:slice-morph}
\end{figure}

A visual representation of the halo morphology assignment is shown in Figure~\ref{fig:slice-morph} for the nDGP2 model. We plot all particles in a 5$\mpcoh$-thick slice of a $30\times30\mpcoh$ portion of one of the three simulation realizations. The positions of each halo are shown as symbols, where stars represent cluster halos and diamonds represent filament halos. In general, there are very few wall or void halos in any simulation, reflecting the hierarchical nature of the cosmic web: even in low-density areas that are voids on large scales, there are smaller scale filaments in which the small halos are embedded (see, e.g., \cite{AragonCalvo:2012bd}). Cluster halos are found at the intersections of the filaments, and most of the halos are identified as being in filaments. Though there are in principle many different ways to classify the cosmic web environment of dark matter halos, the method used here is sufficient to separate halos in crowded cluster environments from the more isolated halos in filaments, and this is reflected in the mass functions.

\begin{figure}[ht]
  \centering{
  \includegraphics[width=15cm]{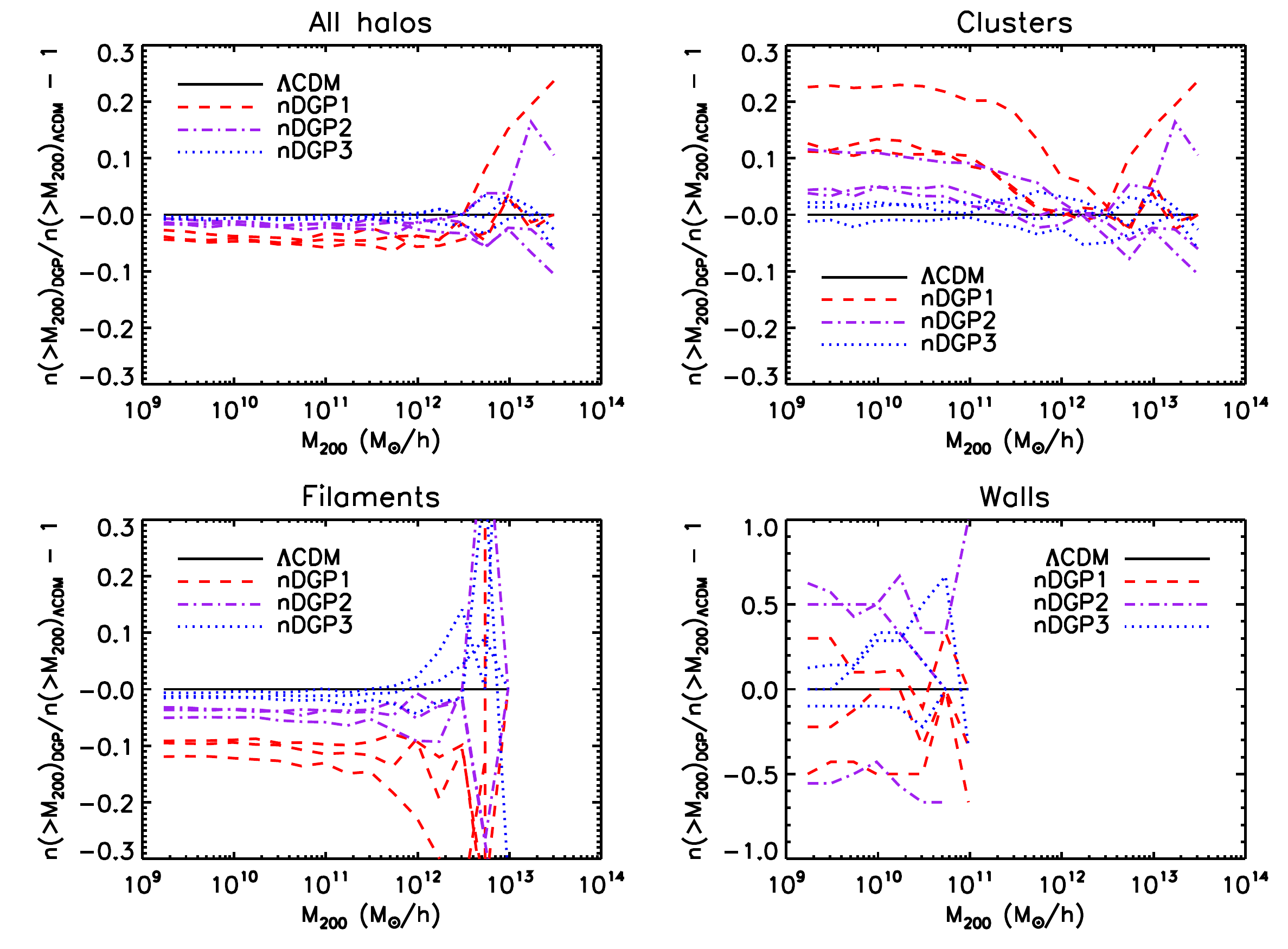}
  }
  \caption{Cumulative mass functions of each nDGP model, shown as the fractional difference between the LCDM simulations, for each of the three independent simulation realizations, for all ORIGAMI halos (top left), cluster halos (top right), filament halos (bottom left), and wall halos (bottom right).
  }
  \label{fig:massfn}
\end{figure}

The halo masses in ORIGAMI are defined as the mass within $R_{200}$, the radius beyond which the density drops below 200 times the critical density, $\rho_{\rm crit}$. Figure~\ref{fig:massfn} shows ratios of the cumulative $M_{200}$ mass functions of the DGP models with respect to the LCDM model. All halos are included in the top left panel, and the other three panels show just cluster, filament, or wall halo mass functions. The mass function ratios for each of the three independent realizations are plotted separately to show the scatter between simulation runs. nDGP3 is very similar to LCDM, only deviating at high masses when the statistics become poor. Being the model with the highest gravity enhancement, nDGP1 deviates the most from the LCDM model. All models begin to deviate at higher masses, where there are very few halos so the statistics are worse; masses above $4\times\,10^{13}\,h^{-1}\,{\rm M}_{\astrosun}$ are dominated by cosmic variance noise and are not shown. 

Note that while it may be unexpected for the nDGP1 mass functions to be below LCDM, this is likely due to the fact that ORIGAMI does not identify subhalos within host halos. There are more massive halos in nDGP1 than in LCDM, and these absorb their subhalos, while for LCDM there will be more distinct density peaks that are eventually identified as individual ORIGAMI halos. There is some difference between the cluster and filament mass functions, especially for nDGP1 which deviates most from LCDM: the deviation from LCDM for low- to mid-mass cluster halos is about 15\% high, while low- to mid-mass filament halos deviate by about 10\% below LCDM. There are very few wall halos, so the wall halo mass function ratios are dominated by noise (and note the different $y$-axis scale). Again, ORIGAMI defines halo morphology by the halo's nearby particles; halos in low-density environments that may be embedded in large scale walls or voids are often surrounded by sub-filaments and thus identified as belonging to a filament. A detailed study of the Vainshtein mechanism in light of the full hierarchical nature of the cosmic web would require both large and high-resolution simulations and is beyond the scope of this paper. 

\begin{figure}[ht]
  \centering{
  \includegraphics[width=12cm]{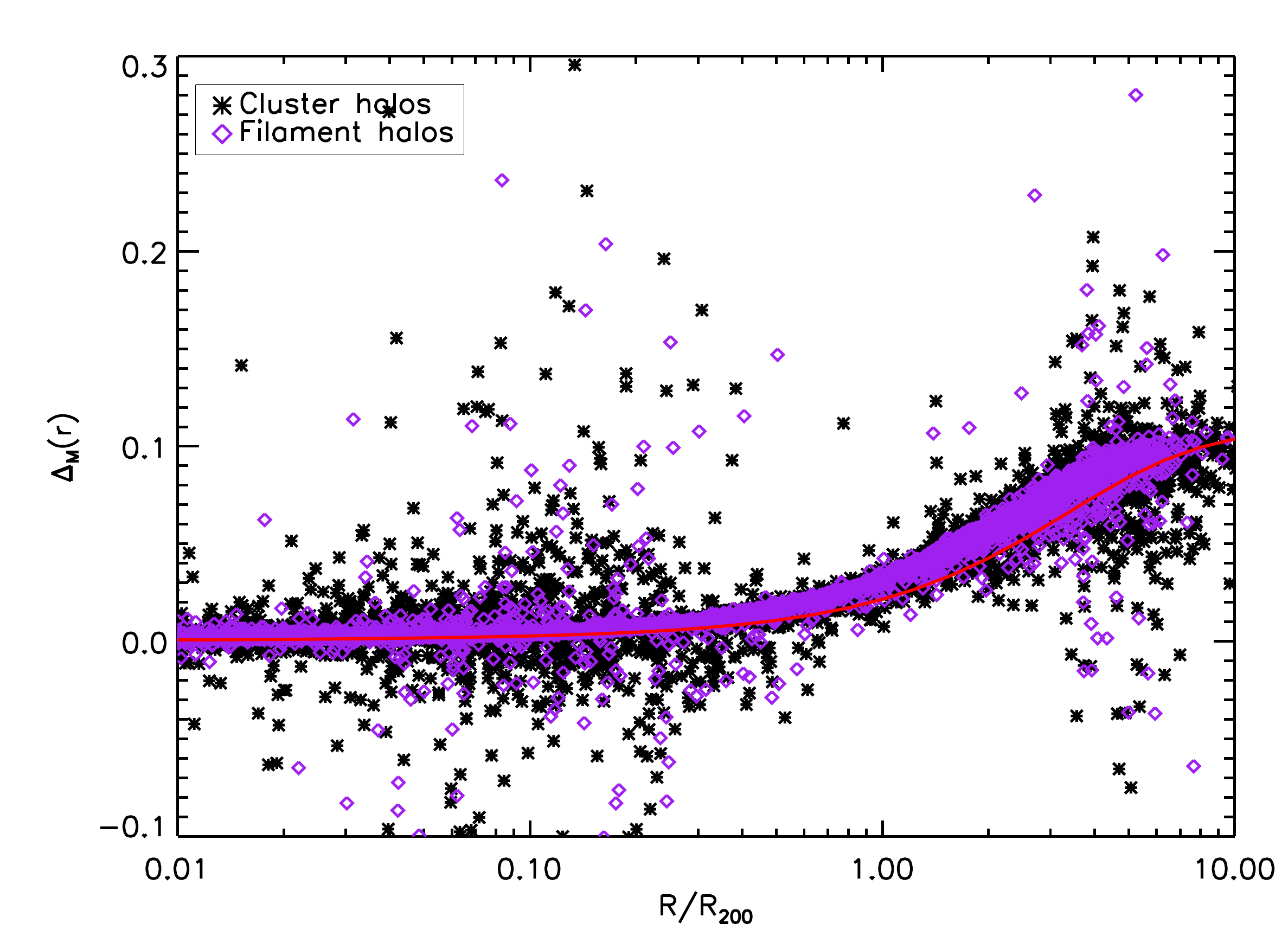}
  }
  \caption{Radial $\Delta_M$ profiles for halos in one realization of the nDGP2 model. Cluster halo profiles are plotted as stars, and filament halo profiles are plotted as purple diamonds; the theoretical prediction from Section~\ref{sec:spherical} is given by the solid red line. Though there is reasonable agreement with theory, there seems to be no dependence of the fifth force enhancement on halo morphology.
  }
  \label{fig:deltam_rad}
\end{figure}

To study the effect of Vainshtein screening, we calculate the ratio between the fifth force and Newtonian force (Equation~\ref{eqn:deltam}) for ORIGAMI halos as a function of normalised radius $R/R_{200}$. The results for the nDGP2 model are shown in Figure~\ref{fig:deltam_rad} for both cluster and filament halos. There is good agreement between the simulation results and the calculation of the spherically symmetric solution from Section~\ref{sec:spherical}, though there is no discernible difference between cluster and filament halos. The fifth force is suppressed until approximately $R_{200}$, and screening becomes less effective in the outer regions of halos as distance from the virial radius increases. This effect is similar, though the magnitude of the fifth force is much reduced, for models closer to LCDM (such as nDGP3). This suggests that probing galactic halos well beyond their virial radii, out to their caustic-crossing or turn-around radii, is more likely to pick up differences between LCDM and models with a Vainshtein screening mechanism (see also~\cite{Lam:2012by, Lam:2013kma, Zu:2013joa, Hellwing:2014nma}).

\begin{figure}[ht]
  \centering{
  \includegraphics[width=12cm]{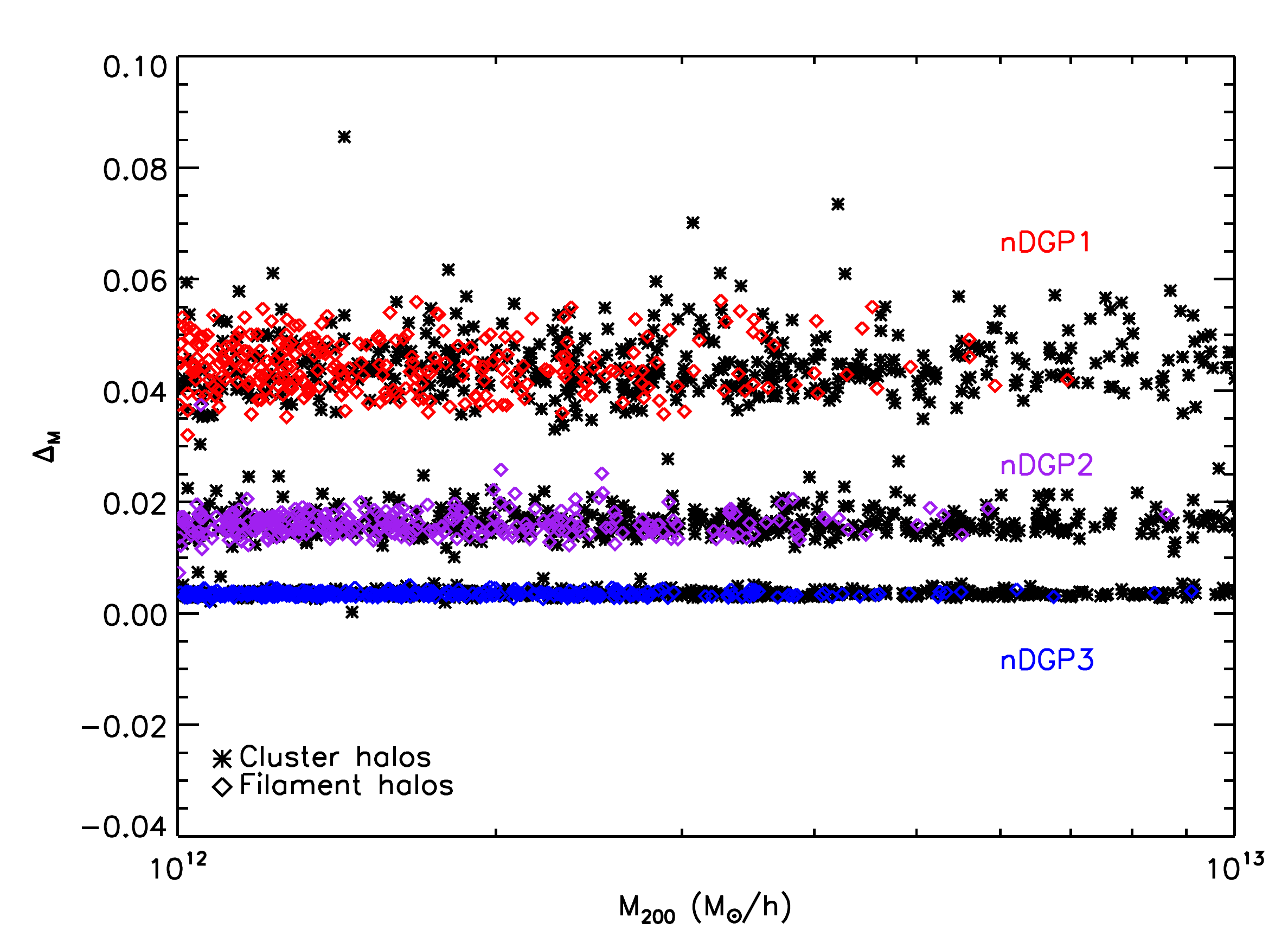}
  }
  \caption{The ratio of the fifth force to Newtonian force, $\Delta_M$, averaged over all radial bins within $R_{200}$ as a function of mass for nDGP1 (top group, red), nDGP2 (middle group, purple), and nDGP3 (lower group, blue), with cluster halos plotted as stars and filament halos as diamonds. Models with a stronger gravity enhancement are further from zero, and there is no significant mass or morphology dependence.
  }
  \label{fig:deltam_mass}
\end{figure}

In addition to $\Delta_M(r)$ profiles, we calculate $\Delta_M$ of the entire halo by averaging the values within $R_{200}$. This is plotted as a function mass for all three nDGP models in Figure~\ref{fig:deltam_mass}, for both cluster and filament halos. As the enhancement to gravity gets smaller, and $r_c$ gets larger, from nDGP1 to nDGP3, the ratio of the fifth force to Newtonian force decreases. There appears to be no dependence of this ratio on either halo mass or morphological environment; the non-dependence on mass was previously confirmed by Ref.~\cite{Schmidt:2010jr}. Since ORIGAMI halos contain particles well outside the virial radius, out to the outer phase-space caustic, we can also obtain $\Delta_M$ for each halo that includes the upturn in the $\Delta_M(r)$ profile in Figure~\ref{fig:deltam_rad}. We find that again there is no dependence on mass or morphological environment, but the $\Delta_M$ per halo tends to increase and the values have larger scatter. This suggests the magnitude of $\Delta_M$ depends on the definition of the virial radius, i.e. the parameter that determines the edge of the halo. For the nDGP models studied here, the value of $\Delta_M$ increases by $\sim 50 - 70$\% when extending the halo from $R_{200}$ to include all ORIGAMI particles, corresponding roughly to the radius of the outer caustic~\cite{Falck}.

\subsection{Velocities}
\label{sec:velocity}
We look at another interesting feature of the Vainshtein mechanism regarding the superimposability of field solutions~\cite{Hu:2009ua}. In the screening mechanisms that rely on non-linearity in the potential or the coupling function to matter, such as the chameleon and symmetron mechanisms, the internal field generated by an object does not superimpose with an external field. Therefore, the field inside the object loses knowledge of any exterior gradient and the fifth force generated by the external field. Thus once the object is screened, it does not feel any fifth force. On the other hand, the non-linear derivative interaction that is responsible for the Vainshtein mechanism enjoys the Galilean symmetry~\cite{Nicolis:2008in}
\begin{equation}
\partial_{i} \phi \to \partial_{i} \phi  + c_i,
\end{equation}
where $c_i$ is a constant vector. This can be easily seen from the fact that the equation of motion for $\phi$ only depends on the second derivatives of $\phi$. If the external fields have wavelengths long compared to the Vainshtein radius, we can regard the gradient of these external fields as constant gradients in the vicinity of an object, and we can always add these constant gradients to the internal field generated by the object. Thus the internal field does superimpose with an external field. This implies that even if the internal field is suppressed by the Vainshtein mechanism, the object still feels the fifth force generated by the external fields~\cite{Hui:2012jb}. 

In order to confirm this theoretical expectation, we study two velocities of dark matter halos. We calculate the velocity dispersion of the halos, measured using particles within $R_{200}$, and the peculiar velocity of the halos, which is just the magnitude of the average velocity of all the particles. The enhancement of velocity dispersions inside the halo with respect to LCDM is suppressed by the Vainshtein mechanism, while the peculiar velocities are enhanced by the linear fifth force induced by the large scale structure. We expect that the ratio of the nDGP velocity dispersion to that of the LCDM halo with the same mass is given by 
\begin{equation}
\frac{\sigma^2_{DGP}}{\sigma^2_{GR}} = 1 + \frac{2}{3 \beta} g \left(\frac{R_{200}}{r_{*}}  \right),
\label{dispersion}
\end{equation}
while the ratio of the nDGP peculiar velocity to that of the LCDM halo with the same mass is given by the linear prediction 
\begin{equation}
\frac{V_{DGP}}{V_{GR}} = 1 + \frac{1}{3 \beta}.
\label{peculiar}
\end{equation}
Here the $\beta$ function defined in Eq.~(\ref{eq:beta}) is evaluated at $z=0$.
 
To find the difference with respect to LCDM, we first need to match halos from the LCDM simulation to corresponding halos in the nDGP simulations. For each LCDM halo having at least 100 particles, we look for the nearest nDGP halo by comparing halo central positions, which in ORIGAMI is the VTFE density-weighted average position of all halo particles. The maximum distance the halos can be from each other is 1$\mpcoh$ in order for there to be a match, and the matched nDGP halo must also have at least 100 particles. While we expect that the nDGP halos should be more massive than the matched LCDM halos, we also require that the ratio between nDGP and LCDM halo masses be within some threshold, which we set to 0.5; using a threshold ratio of 0.1 does not affect the distributions calculated below but results in many fewer matches found.

\begin{figure}[ht]
  \centering{
  \includegraphics[width=15cm]{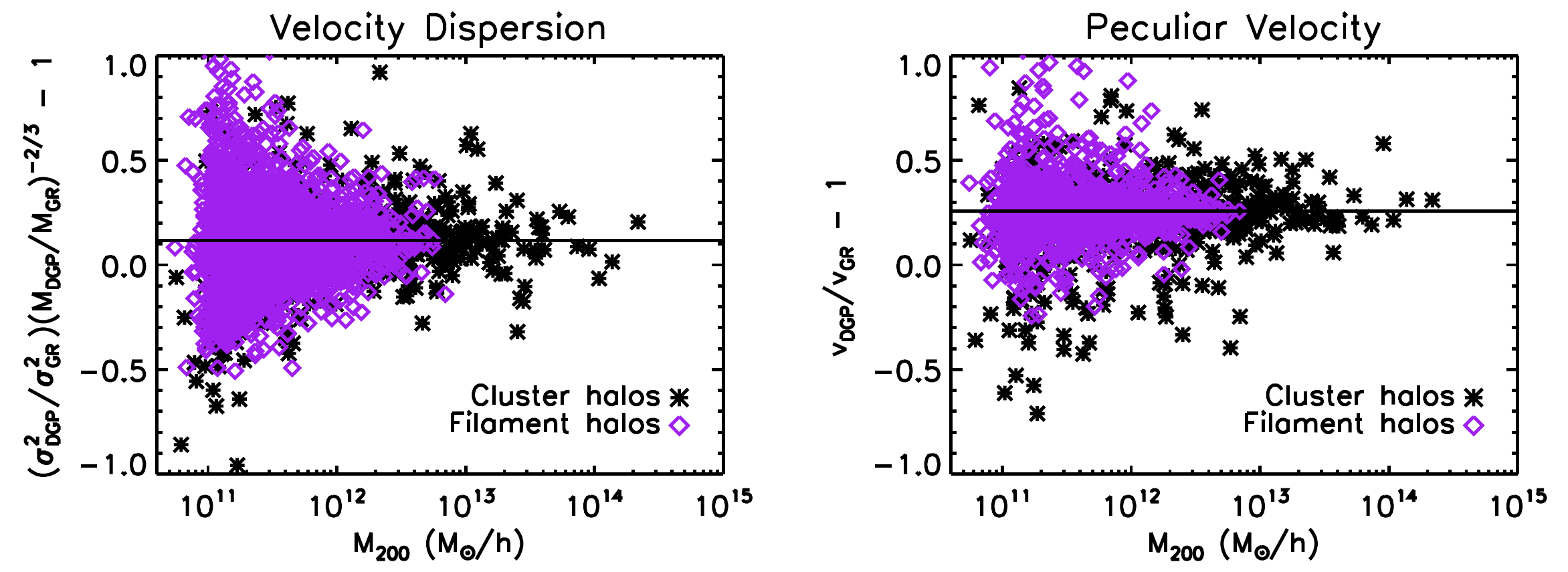}
  }
  \caption{The normalised velocity dispersion (left) and peculiar velocity (right) of ORIGAMI halos in one realization of the nDGP1 simulation, shown as a ratio of the position-matched halos in the LCDM simulation, as a function of halo mass. Cluster halos are plotted as stars and filament halos as diamonds, and the average values for all halos is given by the straight horizontal line.
  }
  \label{fig:velratios}
\end{figure}

The halo velocity dispersion is calculated as the variance of halo particle velocities for those particles within $R_{200}$. 
As in Ref.~\cite{Schmidt:2010jr}, we find no mass dependence of the velocity dispersion after scaling by the virial expectation, $\sigma^2\propto M^{2/3}$. We also find no mass dependence of the ratio of the scaled nDGP velocity dispersion to that of the matched LCDM halos, shown in the left panel of Figure~\ref{fig:velratios} for nDGP1, which has the most deviation from LCDM of the models we simulated. There is again no difference for filament and cluster halos; both show larger scatter at lower masses, and the cluster halos have the highest mass, but the average values of the dispersion ratios are very close for both. The mean of the dispersion ratios for cluster halos is 0.128, and for filament halos it is 0.115, for the simulation realization plotted in Figure~\ref{fig:velratios}. Combining filament and cluster halos, and all three simulation realizations, the mean values of the dispersion are 0.115, 0.0598, and 0.0212 for nDGP1, nDGP2, and nDGP3, respectively. The theoretical prediction of Eq.~(\ref{dispersion}) gives 0.06, 0.02, and 0.002 for nDGP1, nDGP2, and nDGP3, respectively, which agree with the measurements within the large scatters.

The peculiar velocity of the halos is simply the average velocity of all halo particles and reflects the bulk motion of the halo. The ratio of the nDGP to LCDM peculiar velocities of matched halos again shows no mass dependence, shown in the right panel of Figure~\ref{fig:velratios}, and again there is more scatter at lower masses. The cluster and filament halos are again very similar: the mean of the cluster velocity ratios is 0.258, and for filament halos it is 0.255, for the simulation realization plotted. The velocities show a larger enhancement in the nDGP models with respect to LCDM than did the velocity dispersion, as expected. Combining filament and cluster halos and all three simulation realizations, the mean peculiar velocities are 0.253, 0.116, and 0.0292 for nDGP1, nDGP2, and nDGP3, respectively. The linear prediction Eq.~(\ref{peculiar}) gives 0.19, 0.11, and 0.03 for nDGP1, nDGP2, and nDGP3, respectively, which again agrees with the measurements within the large scatters, and the agreement is better than that found for the velocity dispersions. This confirms our picture that the enhancement of peculiar velocities probes the linear fifth force induced by large scale structure, while the enhancement of velocity dispersions within the virial radius is suppressed by the Vainshtein mechanism.

\subsection{Halo Properties}

Here we look at the morphological dependence of the concentrations, spins, and shapes of dark matter halos. Instead of reinventing the wheel, we use the AMIGA Halo Finder (AHF) to calculate these halo properties \cite{Gill:2004km, Knollmann:2009pb}. 
AHF identifies halos by using the spherical overdensity algorithm. It neatly uses the structure of adaptive mesh refinements to find iso-density contours and the centres of potential dark matter halos. It then loops over all particles around a halo centre to remove those which are `gravitationally unbound' by assuming spherical symmetry and comparing the kinetic energy of each particle to the gravitational potential energy at the radius of that particle. After this unbinding procedure, halo properties are calculated using the remaining particles. 

The AHF halo radius and mass are calculated differently than in ORIGAMI. The halo radius is defined to be where the average overdensity of a halo drops below $\Delta_{\rm vir}$ times the mean matter density. For ORIGAMI, this quantity is instead $\Delta = 200$ times the critical density, $\rho_{\rm crit}$. In AHF, $\Delta_{\rm vir}$ depends on the given cosmology and redshift~\cite{Knollmann:2009pb}; here it is the virial overdensity for our base LCDM model. We have not attempted to compute $\Delta_{\rm vir}$ separately for the different nDGP models; the calculation of the quantity involves significant approximations anyway, and the virialisation inside halos is expected to be minimally affected by the fifth force due to the screening.

To determine the morphology of the AHF halos, we match them to ORIGAMI halos using the same method described above: we find the ORIGAMI halo that is closest to the centre of the AHF halo, having at least 100 particles and a mass difference that is within some threshold. This provides a measure of halo morphology from ORIGAMI combined with halo properties calculated with AHF for a set of matched halos. We checked that the AHF mass functions are very similar to the ORIGAMI mass functions in the top left panel of Figure~\ref{fig:massfn}. Since AHF halo masses are defined differently than in ORIGAMI, in the following, to be consistent with the AHF halo properties utilised, we use AHF halo masses.

First we look at the morphology and gravity model dependence of the concentration mass relation. The halo concentration $c=R_{vir}/r_s$ parameterizes the shape of the density profile, where $R_{vir}$ is the virial radius and $r_s$ is the transition point from the inner to outer profile. Typically, the concentration decreases with mass and with redshift \cite{Bullock:1999he, Bhattacharya:2011vr}
in LCDM as well as $f(R)$ simulations \cite{Lombriser:2011zw, Zhang:2013cj}, though the measured concentration depends strongly on whether the halo is relaxed and on the presence of substructure \cite{Neto:2007vq}. For this study we take the AHF concentrations at face value and look for any significant differences as a function of either cosmic web morphology or strength of the Vainshtein mechanism. AHF calculates the halo concentration as defined in Ref.~\cite{Prada:2011jf}.

\begin{figure}[ht]
  \centering{
  \includegraphics[width=15cm]{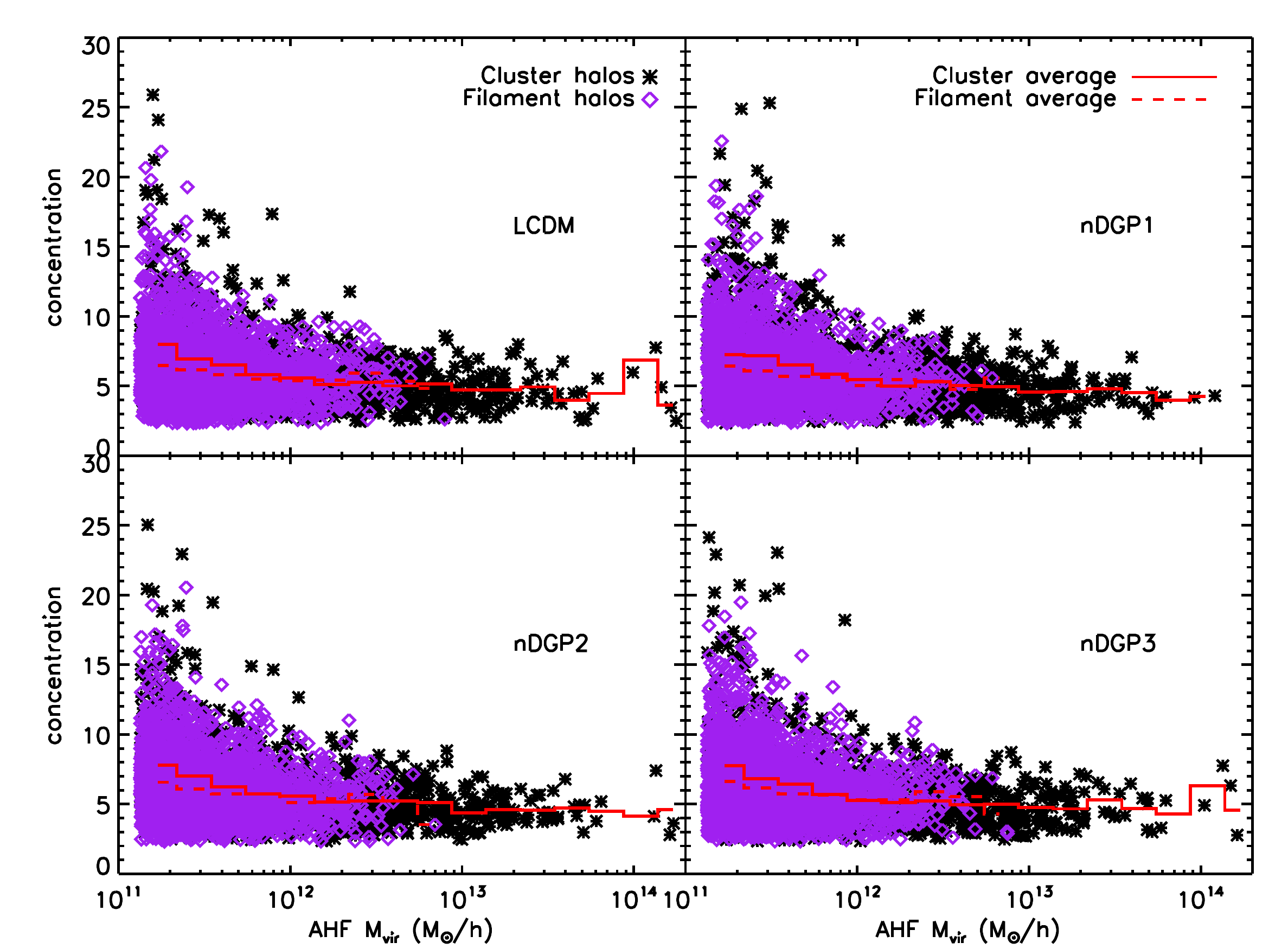}
  }
  \caption{Halo concentration vs. virial mass for one simulation realization of all four gravity models, separated into cluster (stars) and filament (diamonds) halos. The average concentrations, binned in mass, are over-plotted as solid (cluster) and dashed (filament) red lines.
  }
  \label{fig:convmass}
\end{figure}

The concentration versus virial mass of the sample of matched AHF halos is given in Figure~\ref{fig:convmass} for each gravity model and just one simulation realization. Values for each individual cluster and filament halo are plotted along with the mass-binned average values for each halo morphology. For the lowest mass bins, the cluster halos tend to have higher concentrations than the filament halos, which echoes the finding that halos in dense environments tend to be more concentrated than isolated halos \cite{Bullock:1999he}. The mean cluster concentration for nDGP1 is a bit lower than that of LCDM, while nDGP2 and nDGP3 are about the same as LCDM. However, the large scatter prevents us from making definite claims that the nDGP1 halos have lower concentrations than LCDM halos, and any difference between halos of different morphological type are reflected in both LCDM and nDGP simulations.

We turn now to halo spin, which measures the amount of coherent rotation in a system. There has been much attention on measuring the alignment, if any, between the spin of halos within filaments and walls and the orientation of the filaments and walls themselves. This alignment is predicted by the tidal torque theory, by which galaxies acquire their angular momentum from the large scale tidal force field \cite{Peebles:1969jm, White:1984uf}. Tidal torque theory has been investigated in LCDM simulations (see, e.g., \cite{Porciani:2001er, AragonCalvo:2006ay, Hahn:2006mk, Libeskind:2012af, Aragon-Calvo:2013wwa}) and in observations \cite{Tempel:2012xw}, but results so far vary greatly and remain inconclusive. While calculating the spin alignment is beyond the scope of this work, the spin distributions themselves may depend on the cosmic web morphology of halos, though conflicting results have been found so far in LCDM simulations 
\cite{AvilaReese:2005fm, Hahn:2006mk}. 

\begin{figure}[ht]
  \centering{
  \includegraphics[width=15cm]{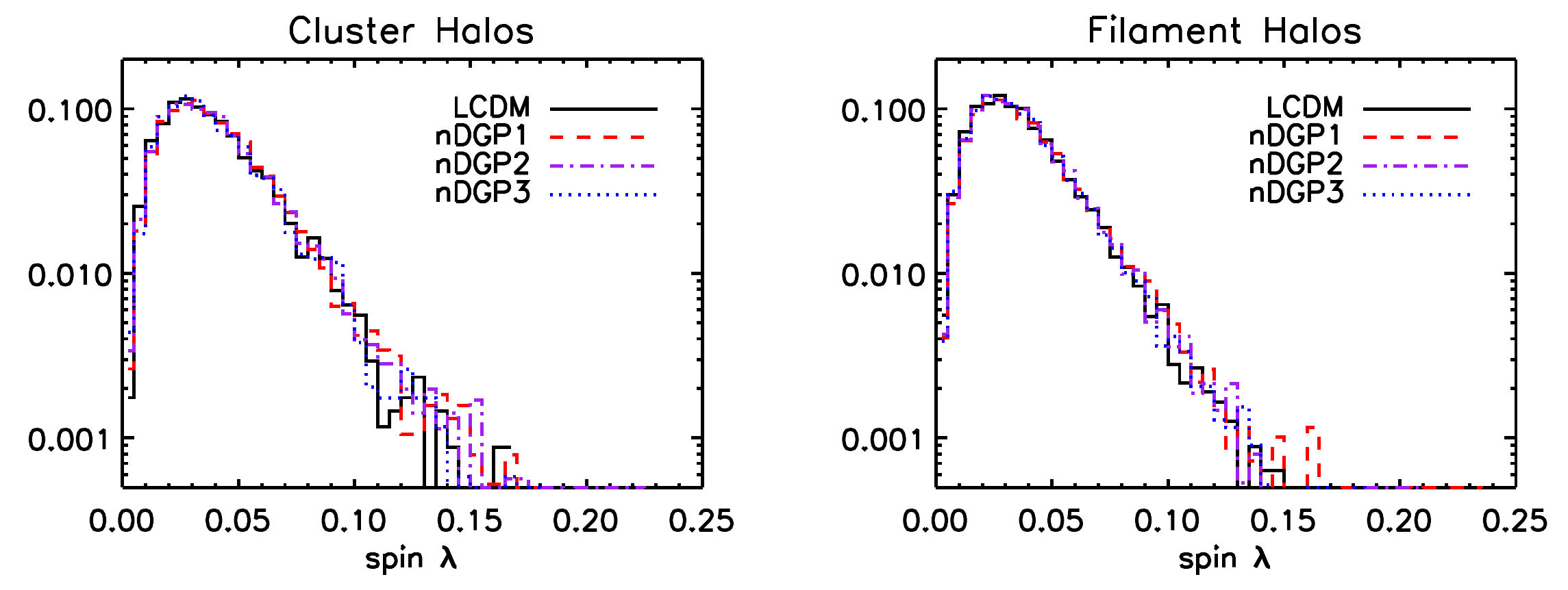}
  }
  \caption{Distributions of halo spins (from all simulation runs) calculated with AHF, for cluster halos (left panel) and filament halos (right panel) for each gravity model. Results from the LCDM simulations are shown as a solid line and the nDGP simulations as dashed lines (see figure legend). There appears to be no dependence of the spin distributions on the gravity model.
  }
  \label{fig:spins}
\end{figure}

Here we calculate the distributions of cluster and filament halo spins to look for any morphology dependence of the Vainshtein mechanism. The AHF code calculates the halo spins as defined in Ref.~\cite{Bullock:2000ry}. We combine the sets of ORIGAMI-matched AHF halos from all three simulation realizations for each gravity model and halo morphology and plot their distributions in Figure~\ref{fig:spins}. It is clear that there is no significant difference between the LCDM and nDGP spin distributions, suggesting again that the morphology dependence of the Vainshtein mechanism is limited to dark matter particles and does not extend to the morphologies of the halos. The distribution of cluster halo spins does seem to have a slightly more skewed tail toward higher spins, which agrees with the tentative findings of Ref.~\cite{Hahn:2006mk} but disagrees with Ref.~\cite{AvilaReese:2005fm}. However, again there is much scatter that prevents us from drawing definite conclusions. The morphology dependence of halo spins in LCDM simulations requires further study which will surely benefit investigations of the effect of modified gravity on halo spins.

Another interesting halo property to study is their shapes. It has been suggested that different symmetron models result in different distributions of halo shapes which can deviate from those of LCDM, while $f(R)$ models show no such difference 
\cite{Llinares:2013jza}. Here we look at the morphology and model dependence of the halo shapes of the matched AHF halos. The halo shapes are quantified by the ratios $b/a$ and $c/a$, where $a$ is the largest axis of the halo's moment of inertia tensor, $b$ is the second largest, and c the smallest; thus $c/a < b/a$, and a spherical halo would have $a = b = c$. The values of the axis ratios are affected both by the particular method of calculation and by the presence of substructure \cite{Knebe:2013xwz}, so here we only concern ourselves with the {\it differences} between axial ratios of halos in different gravity models or with different morphologies.

\begin{figure}[ht]
  \centering{
  \includegraphics[width=12cm]{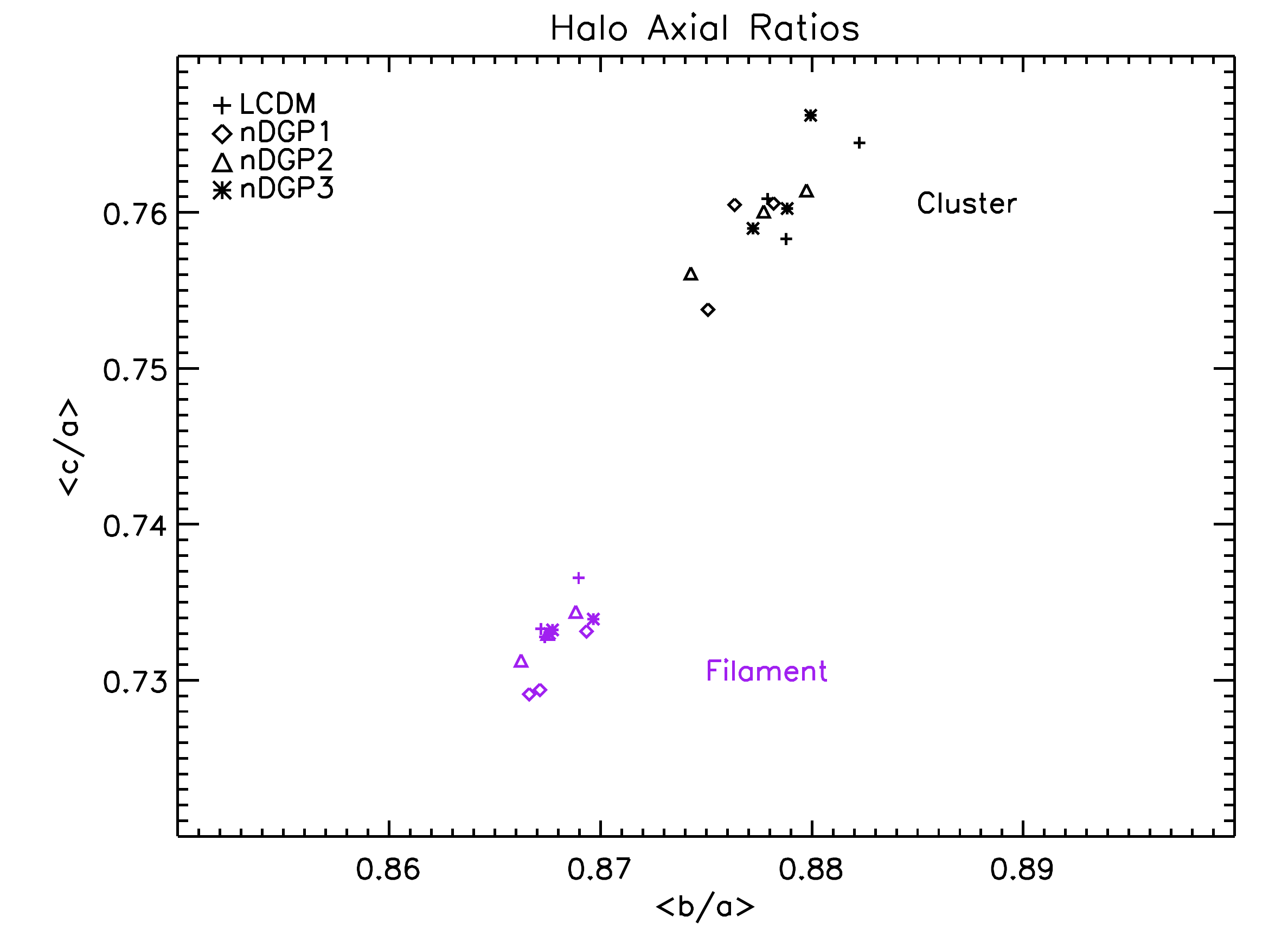}
  }
  \caption{Average halo shapes given by their axial ratios $c/a$ and $b/a$, for each gravity model and simulation run, separated by their ORIGAMI morphology. Note that the distributions of the ratios have a scatter that is larger than the plot range.
  }
  \label{fig:shapes}
\end{figure}

Figure~\ref{fig:shapes} shows the mean axial ratios for the set of matched AHF halos, separated according to their ORIGAMI morphology, for each gravity model. The mean ratios for each of the three individual simulation realizations are plotted to show the variation between simulation runs, and indeed the distributions of the axial ratios for each simulation have a large scatter with a typical standard deviation of $0.07$, which is larger than the plot range.
Though there isn't any significant difference between LCDM and nDGP halo shapes, there is a distinct difference between the average axial ratios of cluster and filament halos. The cluster halos have larger average values of both $c/a$ and $b/a$, making them more spherical than the halos in filaments. This is in disagreement with the finding of Ref.~\cite{Hahn:2006mk} that filament halos are more spherical than those in clusters, but they use a very different definition of halo web environment. Again, note that there is large scatter in the axial ratio distributions (not shown), so this difference is not significant for the current set of simulations.

One thing to point out in this study of halo properties is that AHF halos end at the virial radius, thus all calculated properties depend on the inner regions of the halos likely to experience the full effect of the Vainshtein screening. Given that, any differences between filament and cluster halos would likely be similar to those found in LCDM simulations. However, using particles in the outer regions of halos that are just beginning to spiral into the halo centres will surely increase the difficulty of calculating robust halo properties such as spin and concentration.

For all of these halo properties, very high resolution simulations would help to resolve substructure, distinguish between relaxed vs. unrelaxed halos, and increase the precision with which the properties are calculated. However, given that, as we have shown, halos in Vainshtein models are screened within the virial radius regardless of their mass or web environment, and given that currently, modified gravity simulations are much more expensive to run than LCDM, further investigation of the effect of the cosmic web on halo properties such as spin, concentration, and shape would best be carried out with high resolution LCDM simulations.

\section{Conclusion}

We have studied how the Vainshtein mechanism hides the fifth force in cosmological $N$-body simulations. By identifying the cosmic web morphology of dark matter particles with ORIGAMI, we showed that the Vainshtein mechanism operates effectively to screen halo particles while leaving filament, wall, and void particles unscreened, and this effect is independent of density. Since ORIGAMI morphology is defined according to the dimensionality of the dynamical collapse of individual particles, this means that dark matter particles collapsing along only two or one orthogonal axes feel the full fifth force even if they are in high density environments.  This highlights a key difference between Vainshtein screening and chameleon or symmetron mechanisms.

With the hope of finding an observable signature of the dimensionality dependence of the Vainshtein mechanism, we determined the cosmic web morphology of dark matter halos, which we found to be primarily situated in filaments and clusters. We calculated the ratio of the fifth force to Newtonian force, $\Delta_M$, both as a function of radius from halo centres and as a function of halo mass. Though most halo particles are screened and most filament particles are unscreened, the halos in filaments feel the same fifth force as the halos in clusters. For both cluster and filament halos, $\Delta_M$ is very close to zero until beyond the virial radius and is largest well outside, out to 10 times the virial radius, in good agreement with the theoretical prediction for a NFW halo profile. The average $\Delta_M$ for each halo shows no dependence on either halo mass or cosmic web environment, but it is larger for nDGP models that deviate more strongly from GR.

Another interesting feature of the Vainshtein mechanism is how, in contrast with chameleon and symmetron screening mechanisms, a screened body can still feel the fifth force generated by external fields as long as its wavelength is long compared to the Vainshtein radius. We tested this by measuring the peculiar velocities of dark matter halos and velocity dispersions within the virial radius. We found that, consistent with theoretical expectations, the ratio of nDGP to LCDM peculiar velocities is enhanced by the linear fifth force induced by large scale structure, while the ratio of velocity dispersions is suppressed inside the halo by the Vainshtein mechanism.

Finally, we studied the morphology and gravity model dependence of the concentrations, spins, and shapes of dark matter halos. We found that cluster halos below $\sim 10^{12}\,h^{-1}\,{\rm M}_{\astrosun}$ have slightly higher concentrations than filament halos of the same mass, but this difference is about the same for the LCDM and nDGP simulations. The uncertainties are large, and the morphology dependence of halo concentrations is not well understood even in LCDM. The distributions of halo spins suggest that cluster halos are slightly skewed toward having larger spin, though this effect is small and likely insignificant; additionally we found no significant difference between the spin distributions of LCDM and nDGP halos. We again find no significant difference, and large scatter, between LCDM and nDGP halo shapes; however, the mean values of the axis ratios suggest that cluster halos are slightly more spherical than filament halos. 

In all of these cases, measuring the halo properties is fraught with difficulties, from halo relaxation, to substructure, to methods of calculation, and the dependence of halo properties on their cosmic web morphology is not well understood even for LCDM simulations. When properties are measured in the inner regions of the halos, the screening mechanism is sure to come into effect such that deviations from GR are hard to identify. It is in the outer regions of the halo, beyond the virial radius and out to the caustic or turn-around radius, where modified gravity signatures have a better chance of being detected. Further studies of the morphological dependence of halo properties, both in LCDM and in modified gravity models with the Vainshtein mechanism, would benefit from a larger suite of high resolution, multiple realisation simulations for better statistics.

While it is unfortunate that the morphology dependence of Vainshtein screening for dark matter particles does not appear to hold for dark matter halos, it may be possible to come up with ways of exploiting this feature by tracing along filaments found in the galaxy distribution. Though the galaxies themselves would live in screened dark matter halos, the filament itself may induce a lensing signature (see for example~\cite{Clampitt:2014lna, Higuchi:2014eia}), which may be affected by the enhanced gravity in the filaments in models with the Vainshtein mechanism.  

Another promising area of future work would be the effect of the Vainshtein mechanism on the properties of voids and of halos found in voids, which has been studied so far in $f(R)$ gravity models with the chameleon screening mechanism (see, e.g.,~\cite{Li:2011pj}). It has been found that there are significant differences in the distribution (by volume) of voids in LCDM and $f(R)$ models, and that halos in voids are significantly less screened. Studies of voids in simulations require large box sizes to accurately capture large voids, which are more likely to be observable than small voids, and studies of halos in voids require very high resolution to resolve these usually very small halos. With the types of simulations presented here, we will be able to disentangle the effect of the background expansion from the effect of Vainshtein screening mechanism on the properties of cosmic voids.

The Vainshtein parameters for the simulations in this work were tuned such that they result in the same $\sigma_8$ as specific parameters in the Hu-Sawicki $f(R)$ model. This means that in addition to being able to separate the effects of the cosmological background from the screening, we can directly compare the Vainshtein simulations to simulations of models with the chameleon screening mechanism. The difference between how the Vainshtein and chameleon mechanisms depend on the cosmic web morphology and the environmental density of dark matter particles and halos could provide clues as to how to observationally distinguish between these two different ways of hiding the fifth force in modified gravity models.

\acknowledgments
The authors are grateful for stimulating discussions with Lucas Lombriser, Bhuvnesh Jain, Mark Neyrinck, and Matthew Hull. BF and KK are supported by the UK Science and Technology Facilities Council (STFC) grants ST/K00090/1 and ST/L005573/1. GBZ is supported by the 1000 Young Talents program in China, by the 973 Program grant No. 2013CB837900, NSFC grant No. 11261140641, and CAS grant No. KJZD-EW-T01, and by the University of Portsmouth. BL is supported by the Royal Astronomical Society and Durham University. Numerical computations were done on the Sciama High Performance Compute (HPC) cluster which is supported by the ICG, SEPNet, and the University of Portsmouth.

\end{document}